\documentclass{article}

\usepackage{PRIMEarxiv}

\usepackage[utf8]{inputenc} % allow utf-8 input
\usepackage[T1]{fontenc}    % use 8-bit T1 fonts
\usepackage{hyperref}       % hyperlinks
\usepackage{url}            % simple URL typesetting
\usepackage{booktabs}       % professional-quality tables
\usepackage{amsfonts}       % blackboard math symbols
\usepackage{nicefrac}       % compact symbols for 1/2, etc.
\usepackage{microtype}      % microtypography
\usepackage{lipsum}
\usepackage{fancyhdr}       % header
\usepackage{graphicx}       % graphics
\usepackage{amsmath}
\usepackage{color}
\graphicspath{{media/}}     % organize your images and other figures under media/ folder
\usepackage{multirow}

\title{Robust modestly weighted log-rank tests}

\author{
  Dominic Magirr \\
  Novartis \\
  Basel, Switzerland\\
  \texttt{dominic.magirr@novartis.com} \\
   \And
  Fredrik \"Ohrn \\
  Johnson and Johnson  \\
  Stockholm, Sweden\\
  \texttt{fohrn@its.jnj.com} \\
}

\begin{document}
\maketitle

\begin{abstract}
The introduction of checkpoint inhibitors in immuno-oncology has raised questions about the suitability of the log-rank test as the default primary analysis method in confirmatory studies, particularly when survival curves exhibit non-proportional hazards. The log-rank test, while effective in controlling false positive rates, may lose power in scenarios where survival curves remain similar for extended periods before diverging. To address this, various weighted versions of the log-rank test have been proposed, including the “MaxCombo” test, which combines multiple weighted log-rank statistics to enhance power across a range of alternative hypotheses.

Despite its potential, the MaxCombo test has seen limited adoption, possibly owing to its proneness to produce counterintuitive results in situations where the hazard functions on the two arms cross. In response, the modestly weighted log-rank test was developed to provide a balanced approach, giving greater weight to later event times while avoiding undue influence from early detrimental effects. However, this test also faces limitations, particularly if the possibility of early separation of survival curves cannot be ruled out a priori.

We propose a novel test statistic that integrates the strengths of the standard log-rank test, the modestly weighted log-rank test, and the MaxCombo test. By considering the maximum of the standard log-rank statistic and a modestly weighted log-rank statistic, the new test aims to maintain power under delayed effect scenarios while minimizing power loss, relative to the log-rank test, in worst-case scenarios. Simulation studies and a case study demonstrate the efficiency and robustness of this approach, highlighting its potential as a robust alternative for primary analysis in immuno-oncology trials. 
\end{abstract}

% keywords can be removed

\section{Introduction}

The introduction of checkpoint inhibitors in immuno-oncology has sparked significant debate about whether the log-rank test (or Cox model) should remain the default primary analysis method in confirmatory studies, or if methods more suited to non-proportional hazards should be used instead. The controversy arises because, when these drugs are compared to chemotherapy, survival curves often stay similar for several months before diverging \cite{rahman2019deviation, castanon2020critical}. Although the log-rank test is effective in controlling the false positive rate, it may lose power in this scenario.

Several proposals have suggested replacing the log-rank test with a weighted version designed to detect long-term survival improvements. Since the log-rank test works by taking a sum across event times of the observed minus the expected (under the null) number of events on the experimental arm, it is conceptually straightforward to give larger weights to the later event times relative to earlier event times. There are, of course, an infinite number of possible ways to do this, varying in the extent to which they upweight the later data. Uncertainty at the design stage regarding the extent of delayed effect can make it a risky business to pre-specify a particular weight function. For this reason, further proposals have been made to use a test statistic that is the maximum of multiple weighted log-rank statistics, each with a different weight function \cite{karrison2016versatile, lee2007versatility, ristl2021delayed, tarone1981distribution, wang2021simulation, ghosh2022robust}. Such test statistics have reasonable power across a large range of possible alternative hypotheses. In particular, a cross-pharma working group recently recommended the "MaxCombo" test  \cite{roychoudhury2021robust} which combines several statistics from the Fleming-Harrington family of weight functions \cite{fleming2011counting}. 

Despite some attractive properties of the MaxCombo test, uptake has been slow. The method has been criticized for occasionally producing counterintuitive results. For example, Freidlin and Korn \cite{freidlin2019methods} construct a scenario in which the experimental treatment is uniformly worse than control, and yet the MaxCombo test would have a high chance of rejecting the null hypothesis in favor of the experimental treatment. To address this specific issue, Magirr \& Burman developed the modestly weighted log-rank test \cite{magirr2019modestly}. It still gives larger weight to later event times than earlier event times, making it suitable for delayed effect scenarios, but the relative weighting is carefully controlled, such that test statistic does not reward a detrimental effect of the experimental treatment at early timepoints.

The modestly-weighted test was introduced as a single test statistic. Its constraint on the relative weighting provides a reasonable level of robust power for proportional hazards scenarios, as well as different degrees of delayed effect \cite{magirr2019modestly, magirr2021non}. If, however, contrary to expectation, there is an early effect of treatment that diminishes over time, then the modestly-weighted test has considerable power loss compared to a log-rank test \cite{magirr2021non}. This lack of robustness in a worse-case scenario perhaps makes the test less attractive to practitioners, especially given the status of the log-rank test as a de-facto standard for the primary analysis.

In this paper, we make the concrete proposal to use a test statistic that combines the best features of the standard log-rank test, the modestly-weighted log-rank test, and the MaxCombo test. We build on the work of Ghosh et al.\cite{ghosh2022robust}, who investigated a test statistic based on the maximum of two modestly-weighted log-rank statistics. We, instead, consider the test statistic based on the maximum of the standard log-rank statistic and one modestly-weighted log-rank statistic.  The component test statistics can be very strongly correlated under the null hypothesis. For this reason, both components can be assessed with minimal $\alpha$ adjustment, meaning that even under worse-case scenarios there is minimal power loss compared to the log-rank test. On the other hand, gains in power can be large under delayed effect scenarios. We demonstrate the desirable properties of the test via a simulation study and a case study. 

%Delayed response is a reality, in particular immuno-oncology. Power loss can be considerable, how to deal with the issue? 

%Add previous publications. Need to comment on papers about MaxCombo, in what sense is what we are doing different or novel? 

%\begin{itemize}
%\item Issues with MaxCombo articulated by Magirr and Burman
%\item Correlation as high? What we are proposing could have lower critical value (closer to nominal $\alpha$, will have to see how this translates into power/assurance gains
%\end{itemize}

%Power can do down quickly if we have a delay that is not accounted for. On the other hand, planning for a test that would be beneficial in case there is a delay, can be less powerful than the standard test if there is a no delay. For a dual test, that considers both the standard log rank test and mWLRT, the null hypotheses can be strongly correlated. For this reason, both components can be assessed with minimal $\alpha$ adjustment. On the other hand, considering power and assurance metrics, gains can be quite large and in particular the approach can be more robust to deviations from assumptions. 

%What is the scope? Consider single PFS analysis or more complex (GSD, OS, etc). Futility stopping? Discussed May 14, think we can illustrate many of the main points focssing on a single PFS analysis. 

\section{Methodology applied to an example data set}

\subsection{Weighted log-rank test}

Following the notation in Ghosh et al. \cite{ghosh2022robust}, a generic weighted log-rank statistic is defined as
\begin{equation}
    G^W = \sum_{i = 1}^k W(\tau_l) \left[ d_{1l}- E(d_{1l})\right],
\end{equation}
where $d_{j1},\ldots,d_{jk}$ correspond to the number of events occurring at event times $\tau_1 < \ldots < \tau_k$ on treatment $j = 0,1$,  and $d_{l} = d_{0l} + d_{1l}$ for $l= 1,\ldots,k$. The expectation is taken under the assumption of identical survival curves, 
\begin{equation*}
    E(d_{1l})  = \frac{n_{1l}d_{1l}}{n_l},
\end{equation*}
as is the variance,
\begin{equation*}
    \mathrm{var}(d_{1l}) = \frac{n_{1l}(n_l - n_{1l})d_l(n_l-d_l)}{n_l^2(n_l - 1)},
\end{equation*}
where $n_{jl}$ and $n_l$ represent the numbers at risk at event time $\tau_l$ on treatment $j$ and in total, respectively, with 
\begin{equation*}
    \mathrm{var}(G^W) = \sum_{l=1}^k  \left[ W(\tau_l) \right]^2 \mathrm{var}(d_{1l}).
\end{equation*}
The weight $W(\tau_l)$ associated with event time $\tau_l$ can be chosen in a variety of ways. We focus on two classes of weight function. Firstly, \textit{modestly weighted log rank tests} \cite{magirr2019modestly} 
\begin{equation}
    W^{s^*}(\tau_l) = \frac{1}{\max \left\lbrace \hat{S}(\tau_l-), s^* \right\rbrace}
\end{equation}
where $\hat{S}(\tau_l-)$ is the Kaplan-Meier estimate of survival just prior to event time $\tau_l$ based on the pooled data. The first event time is given a weight of $1$, whereafter weights increase gradually up to $1 / s^*$, whereafter they remain constant. The parameter $s^*$ can be used to fine tune the test. We focus on $s^*=0.5$, so that weights range from $1$ to $2$.

The second class of weight function we consider is the Fleming-Harrington ($\rho$, $\gamma$)  family \cite{fleming2011counting} where
\begin{equation}
    W^{\rho, \gamma}(\tau_l) = \hat{S}(\tau_l-)^\rho \left\lbrace  1 - \hat{S}(\tau_l-) \right\rbrace ^ \gamma.
\end{equation}
We focus on $(\rho, \gamma)=(0, 0.5)$, so that the first event is given a weight of $0$, whereafter weights gradually increase from $0$ towards $1$.

 Under a null assumption of equal survival curves on treatments $0$ and $1$,  the weighted log-rank statistic,
 \begin{equation*}
      Z^{W} = \frac{G^W}{\sqrt{\mathrm{var}(G^W)}}
 \end{equation*}
asymptotically follows a standard normal distribution.

\subsection{Example}

\begin{figure}
\centering
\includegraphics[width=0.6\linewidth]{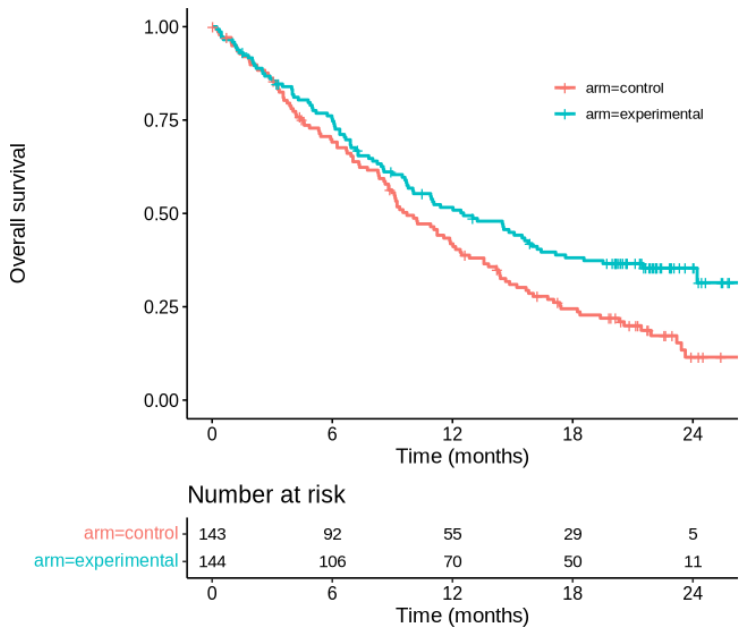}
\caption{\label{figPoplar} Visualization of the POPLAR data set.}
\end{figure}

Figure\ref{figPoplar} displays the results of the POPLAR trial \cite{fehrenbacher2016atezolizumab, gandara2018blood}, where an experimental immuno-oncology agent was compared with a control arm in terms of overall survival. The one-sided p-values from three different weighted log-rank tests are presented in Table \ref{tabResultsPoplar}. Compared to the standard log-rank test ($p = 0.0028$), both the modestly weighted log rank test ($s^*=0.5)$ and the Fleming-Harrington-($0,0.5)$ test result in lower p-values ($p = 0.0009$ and $p=0.0006$, respectively).

\begin{table}
\centering
\begin{tabular}{ |c c | c | c |}
\hline 
\multicolumn{2}{|c|}{Test}& $\alpha$ split & One-sided p-value \\\hline \hline
\multirow{4}{*}{rMW} & LR & (1, 0) &  $0.0028$ \\
& & (0, 1) & $0.0009$ \\
&\multirow{2}{*}{} & (0.5, 0.5) & $0.0012$  \\
& & (0.6, 0.4) & $0.0015$ \\  
\hline
\multirow{4}{*}{Max Combo} & LR & (1, 0) &  $0.0028$ \\
&  & (0, 1) & $0.0006$ \\
&\multirow{2}{*}{} & (0.5, 0.5) & $0.0009$  \\
& & (0.6, 0.4) & $0.0012$ \\  
\hline
\end{tabular}
\caption{\label{tabResultsPoplar} Application of weighted tests to the POPLAR data set. The robust Modestly Weighted (rMW) test uses a combination of a standard log-rank (LR) test and a modestly-weighted test with $s^*=0.5$. The Max Combo test uses a combination of a standard log-rank test and a Fleming-Harrington-(0, 0.5) test. When the full alpha is allocated to the LR component of the test then the test reduces to the standard LR test. When the full alpha is allocated to the other component of the test then the test reduces to either the modestly-weighted test or the Fleming-Harrington test.}
\end{table}

\subsection{MaxCombo type tests}
%{\fredrik{consider name "combination test" given use in adaptive design}}

Instead of relying on a single weighted log-rank test, it has also been suggested to use a ``max-combo'' type statistic \cite{mukhopadhyay2022log, roychoudhury2021robust, royston2020simulation, ghosh2022robust}, which is defined in general as
\begin{equation}
    T^W = \max(Z^{W_1}, Z^{W_2})
\end{equation}
for suitable choice of weights $W_1$ and $W_2$. As described in Ghosh et al. \cite{ghosh2022robust}, under a null assumption of equal survival curves on treatments $0$ and $1$,  the weighted log-rank statistics $Z^{W_1}$ and $Z^{W_2}$ asymptotically follow a joint normal distribution with
$$\mathrm{E}(Z^{W_1}) = \mathrm{E}(Z^{W_2}) = 0$$
$$\mathrm{var}(Z^{W_1}) = \mathrm{var}(Z^{W_2}) = 1$$
\begin{equation}\label{eqCov}
  \mathrm{cov}(Z^{W_1}, Z^{W_2}) = \frac{\sum_{l=1}^{k_j}W_1(\tau_l)W_2(\tau_l)\mathrm{var}(d_{1l})}{\sqrt{\sum_{l=1}^{k_j} \{\left[  W_1(\tau_l) \right]^2 \mathrm{var}(d_{1l}) \} \sum_{l=1}^{k_j} \{ \left[  W_2(\tau_l) \right]^2 \mathrm{var}(d_{1l}) \} }}.  
\end{equation}
When the component tests come from the modestly-weighted family, we will refer to these MaxCombo type tests as robust modestly weighted log rank tests (rMW), as they are designed to be robust to deviations from assumptions made at the design stage, and to distinguish them from the original MaxCombo family of tests that use Fleming-Harrington weight functions in the component tests.

Taking the POPLAR data set in Figure \ref{figPoplar} as an example, when $W_1 \equiv 1$  (a standard log-rank test) and $W_2$ is a modestly weighted log-rank test ($s^*=0.5$), then the estimated correlation  between $Z^{W_1}$ and  $Z^{W_2}$ is approximately $0.97$.

The joint null distribution can be used to find a critical value for performing a test of the null hypothesis. In the simplest case that we wish to split $\alpha$ equally to the two test statistics we find $c$ such that 
\begin{equation}
    P(T^W > c) = \alpha.
\end{equation}
As a variation, we may wish to assign more weight to one of the component test statistics (say, $Z^{W_1}$) compared to the other. One way to achieve this would be to pick $k_1$ and $k_2$, where $0 \leq k_2 < k_1 \leq 1$ and $k_1 + k_2 = 1$, and find $c^{\prime}$ such that
\begin{equation}
    P(\left\lbrace Z^{W_1} > c^{\prime}\Phi^{-1}(1 - k_1 \alpha) \right\rbrace \cup \left\lbrace Z^{W_2} > c^{\prime}\Phi^{-1}(1 - k_2 \alpha)  \right\rbrace) = \alpha.
\end{equation}

Based on the POPLAR data set, and a combination of a log-rank test and a modestly weighted log-rank test ($s^*=0.5$), with an equal splitting of $\alpha = 0.025$ between the two tests the critical value is $c = 2.04$. If, on the other hand, we prefer to give more weight to the standard log-rank test via $(k_1, k_2) = (0.6, 0.4)$ then the critical value for the standardized log-rank statistic would be $c^{\prime}\Phi^{-1}(1 - k_1 \alpha) = 1.99$, while the critical value for the standardized modestly-weighted log-rank statistic would be $c^{\prime}\Phi^{-1}(1 - k_2 \alpha) =2.13$. A one-sided p-value can be found by searching for an $\alpha$ whereby the test statistic matches the critical value exactly. As shown in Table \ref{tabResultsPoplar}, for the equal alpha splitting this p-value is $0.0012$, while for the case $(k_1, k_2) = (0.6, 0.4)$ the p-value is $0.0015$. Also, shown in Table \ref{tabResultsPoplar} are the p-values from the combination test using the standard log-rank test together with the Fleming-Harrington-($0,0.5$) test. In this case the correlation between the test statistics is slightly lower ($0.94$) resulting in a critical value of $2.08$ when an equal alpha-split is used.
%Some simple graphs illustrations of the data, motivating why we think our approach could have benefits. 

%Report simple high level results for the Roche data using different approaches. We can see that eg mWLRT would have been beneficial, but is it robust? 

%Write down what we are proposing in mathematical text

%Could start with simplest possible case, eg one PFS analysis in a single full population. Think about whether we'd also like to consider GSD, testing of OS? With more than one endpoint, could make the case that we'd be allowed to test the secondary endpoint (OS) at full $alpha$ even if just one "component" shows an effect. 

\section{Evaluation of operating characteristics}

We use a simulation study to investigate type 1 error control and power under a range of assumptions.

\subsection{Scenarios}

\begin{figure}
\centering
\includegraphics[width=\linewidth]{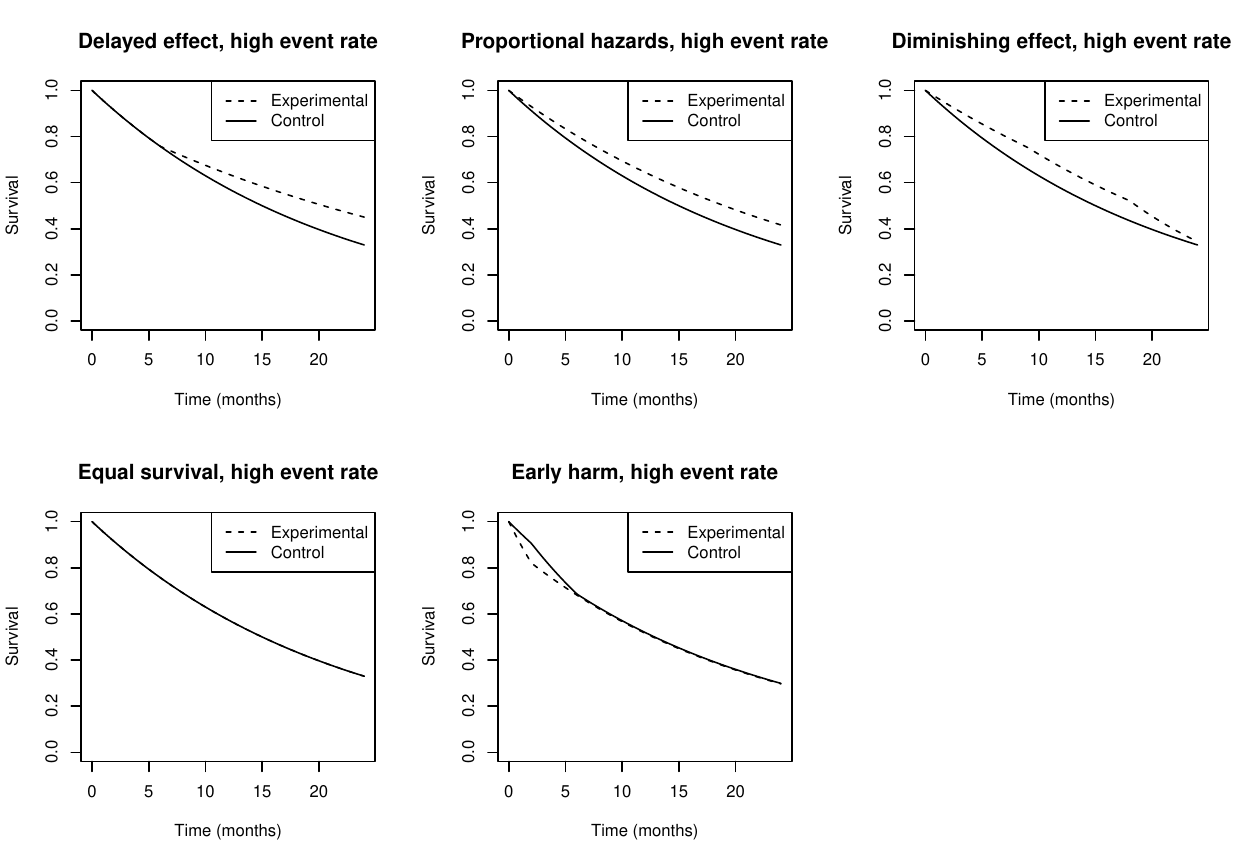}
\caption{\label{figHigh} Scenarios from an immuno-oncology setting with high data maturity.}
\end{figure}

\begin{figure}
\centering
\includegraphics[width=\linewidth]{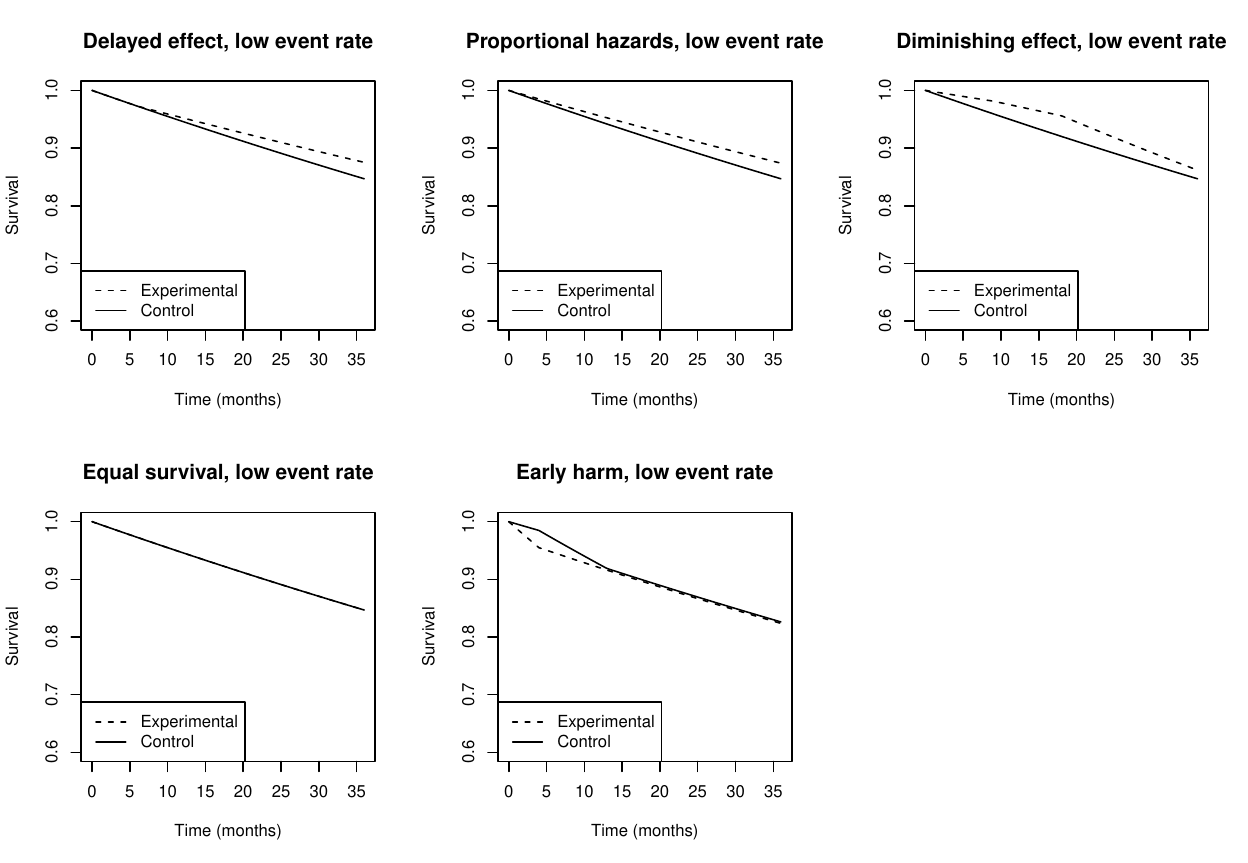}
\caption{\label{figLow} Scenarios from a cardiovascular setting with low data maturity.}
\end{figure}

We consider two sets of scenarios in Table \ref{tabScenarios}. The "high event rate" scenarios (Figure \ref{figHigh}) are designed to mimic a typical oncology trial in aggressive disease, where the majority of participants experience the event of interest. The "low event rate" scenarios (Figure \ref{figLow}) are based on large-scale cardiovascular outcome trials with large sample sizes but where only a small proportion experience the event of interest. Within each set of scenarios, we consider the power under one case of delayed effect, one case of proportional hazards, and one case of diminishing effect. In addition, we consider the one-sided type 1 error rate, both under an equal survival scenario and a situation where the experimental arm is uniformly worse than the control arm. We use piecewise exponential event distributions. The recruitment rate is uniform over 12 months. The only source of censoring is the administrative censoring at the end of the study. In our case, we fix the total study length. The sample sizes have been chosen such that the log-rank test has approximately 80\% power.

\begin{table}
\centering
\begin{tabular}{ |l l | c | c | l l | l l |}
\hline
\multicolumn{2}{|c|}{\multirow{2}{*}{Scenario}} & Study & \multirow{ 2}{*}{N} & \multicolumn{2}{|l|}{Experimental arm} & \multicolumn{2}{|l|}{Control arm}\\ 
& & length  & & Knots & Hazard rates  &  Knots & Hazard rates  \\\hline \hline
 & Delayed effect & \multirow{ 5}{*}{24}&\multirow{ 5}{*}{1000} &  $6$  & $0.0462, 0.0289$ & $\cdot$  & $0.0462$ \\
High&Proportional hazards & & & $\cdot$ & $0.0365$ & $\cdot$ & $0.0462$  \\
event&Diminishing effect & & & $9,18$ & $0.0315,0.0408,0.0693$ & $\cdot$ & $0.0462$   \\
rate&Equal survival&  & &  $.$ & $0.0462$  & $\cdot$ & $0.0462$   \\  
&Early harm&  & & $2$ & $0.0990,0.0462$ & $2,6$ & $0.0495, 0.0693, 0.0462$  \\
\hline
\hline
& Delayed effect& \multirow{ 5}{*}{36} & \multirow{ 5}{*}{6000}& $6$ & $0.00462,0.00352$ & $\cdot$  & $0.00462$ \\
Low&Proportional hazards& & & $\cdot$ & $0.00375$ & $\cdot$  & $0.00462$\\
event&Diminishing effect& & & $9,18$ & $0.00210,0.00289,0.00578$ & $\cdot$  & $0.00462$  \\
rate&Equal survival& & & $\cdot$ & $0.00462$ & $\cdot$  & $0.00462$  \\  
&Early harm& & & $4$ & $0.01160,0.00462$ & $4,13$ & $0.00385,0.00770,0.00462$ \\
\hline
\end{tabular}
\caption{\label{tabScenarios} Simulation scenarios. In all scenarios, patients are recruited at a uniform rate over 12 months. }
\end{table}

\subsection{Methods}

We compare the following six methods.

\begin{enumerate}
    \item Standard log-rank test (LR).
    \item Modestly-weighted log-rank test (MW) with $s^*=0.5$.
    \item Robust modestly-weighted log-rank test (rMW) combining a standard log-rank test with a MW test with $s^* = 0.5$, and where $k_1 = k_2 = 0.5$. 
    \item Robust modestly-weighted log-rank test (rMW) combining a standard log-rank test with a MW test with $s^* = 0.5$, and where $k_1 = 0.6, k_2 = 0.4$.
    \item Fleming-Harrington-(0, 0.5) test (FH).
    \item MaxCombo test combining a standard log-rank test with a FH(0, 0.5) test, and and where $k_1 = k_2 = 0.5$. 
\end{enumerate}

\subsection{Implementation}

For each simulated data set, we calculate the weights for the component tests using the nphRCT R package \cite{nphRCT}. Subsequently, we use (\ref{eqCov}) to estimate the joint null distribution of the component tests and thus derive a cut-off value for the maximum Z statistics to give a one-sided type one error rate of $\alpha = 0.025$. We then compare the cut-off value with the component test statistics, which are also available via nphRCT. Code to reproduce results is available at \url{https://github.com/dominicmagirr/modestMaxCombo/}.

\subsection{Results}

Simulation results are presented in Table \ref{tabResultsHigh}. All methods control the one-sided type 1 error rate at $\alpha = 0.025$ under the scenario of equal survival curves on the two treatment arms. However, for the scenario with early harm but where the experimental arm is uniformly worse than the control arm, both the Fleming-Harrington-(0, 0.5) test and the MaxCombo test reject the null hypothesis in favour of the experimental arm more than 2.5\% of the time. In particular for the low event rate scenarios, the inflation is large.

Under the delayed effect scenarios, all methods improve on the power of the standard log-rank test. For the high event rate scenarios, the improvement is substantial for all methods, with the Fleming-Harrington-(0,0.5) test having the largest improvement. For the low event rate, only the FH test and MaxCombo test offer a substantial improvement but at the cost of considerable type I error inflation.

As expected, the log-rank test is the most powerful under the proportional hazards assumption, but the competing methods do not lose a tremendous amount of power, especially the robust tests. For the diminishing effect scenario with a high event rate, the power loss of the modestly-weighted and FH tests compared to the LR test is strikingly large. However, this is largely recovered in the robust versions of the tests.

Altering the parameter $k_1$ from $0.5$ to $0.6$ to give more weight to the LR test appears to have a very small effect on the operating characteristics. It demonstrates however, that it is possible to keep any power loss compared to the LR test at a negligible level, whilst still offering a meaningful improvement in power under delayed effect scenarios. This is possible due to the high correlation between the component test statistics under the null hypothesis, while it is still possible to achieve different outcomes under the alternative. Getting very close to the LR test (power of 0.77 is reported for both LR and rMW(k=0.6) for the high event rate proportional hazards scenario) is likely to be important for getting the rMW to be used in practice, since most trials are still designed for the LR test under an assumption of proportional hazards. 

Another way to assess the robustness of the six tests across scenarios would be to consider the assurance \cite{o2005assurance}. Based on the prior belief about the scenarios for survival benefit at the design stage, we can summarize the power metrics across scenarios for a given test in Table \ref{tabResultsHigh} into a single number. For illustration purposes we apply a discrete prior for the different scenarios considered and refer to \cite{salsbury2024assurance} for a more detailed discussion of assurance methods to design trials with non-proportional hazards. We focus on the high event rate case, placing prior probability of 1/3 on the delayed effect, proportional hazards and diminishing effect scenarios. The assurance for the two versions of rMW as well as for the MaxCombo test is between 0.78-0.79, 0.77 for the standard LR, 0.73 for MW and 0.70 for FH. These numbers illustrate the robustness of rMW across the scenarios considered. MaxCombo also performs well with respect to assurance, but suffers from type I error inflation under the early harm scenario in Table \ref{tabResultsHigh}. LR, MW and FH provide high power under specific scenarios, but the two latter in particular do not perform as well for an assurance metric that averages power across scenarios. If there is a high prior probability of proportional hazards it would be very reasonable to stick with LR as the primary analysis, since when the prior probability of proportional hazards tends to 1 we know that LR is the most powerful test which would also be reflected in the assurance metric. When there is uncertainty about the survival benefit at the design stage however, the two rMW tests presented offer appealing operating characteristics with few downsides. 

%> asr.high.equal3
%  asr.LR    asr.MW asr.rMW.equal asr.rMW.unequal asr.FH asr.MaxCombo
%1   0.77 0.7333333     0.7833333       0.7866667    0.7    0.7866667

%Consider adding average power, and/or graph with alpha split on the x axis?

%{\fredrik{Assurance quick and dirty for prior with 1/3 on delayed/proportional/Diminishing in parenthesis)
%}}

\begin{table}
\label{table_assumptions}
\centering
\begin{tabular}{ |c  c | c c c  c c c |}
\hline
\multicolumn{2}{|c|}{\multirow{3}{*}{Scenario}}  & \multicolumn{6}{|c|}{Power}\\ 
& & LR & MW  & rMW & rMW & FH  & MaxCombo\\
& & & & $k_1=0.5$ & $k_1=0.6$ & & \\
\hline \hline
\multirow{ 5}{*}{High event rate} & Delayed effect &   0.79 & 0.88 & 0.87 & 0.85 &  0.92 & 0.90 \\
&Proportional hazards                              &   0.77 & 0.75 & 0.76 & 0.77 & 0.72 & 0.75 \\
&Diminishing effect                                &  0.75 & 0.57 & 0.72 & 0.74 & 0.46 & 0.71 \\
&Equal survival                                    & 0.024 & 0.024& 0.024& 0.025& 0.025& 0.025 \\  
&Early harm                                        & 0.007 & 0.021& 0.015& 0.012& 0.056& 0.044 \\
\hline
\hline
\multirow{ 5}{*}{Low event rate} & Delayed effect & 0.79 & 0.80 & 0.80 & 0.79 &  0.86 & 0.84 \\
&Proportional hazards                             & 0.79 & 0.79 & 0.79 & 0.79 & 0.74 & 0.78 \\
&Diminishing effect                               & 0.79 & 0.73 & 0.79 & 0.79 & 0.14 & 0.76 \\
&Equal survival                                   &0.024 & 0.024& 0.024& 0.024& 0.024& 0.025 \\  
&Early harm                                       &0.009 & 0.013& 0.01 & 0.009& 0.154& 0.127 \\
\hline
\end{tabular}
\caption{\label{tabResultsHigh} Simulation results.}
\end{table}

\section{Summary and conclusions}

We have demonstrated that rMW provides robust power across a range of assumptions, including a delayed effect, proportional hazards and a diminishing effect. The results in Table~\ref{tabResultsHigh} show that while allocating all $\alpha$ to one test can achieve power gains when evaluated under a specific scenario, the rMW can achieve adequate power across the range of scenarios. Some of the other tests considered suffer if the prior assumption that motivated the choice of test is not met. Hence we believe that when there is uncertainty about the nature of the survival benefit at the design stage, the probability of a positive read-out can be increased or made more robust by taking the different possible scenarios into account. In particular, the $k$ parameter in the rMW test can be tuned to get very close to the power of the standard LR under proportional hazards with meaningful power gains under a delayed effect scenario.  

The rMW makes it possible to assess various test statistics simultaneously, making use of the fact that they are strongly correlated under the null. However, the survival benefit under the alternative differs across the scenarios which makes it possible to have diverging results depending on the choice of test. The robustness of the rMW and other candidate tests can also be assessed with an assurance metric obtained by placing prior probabilities on possible scenarios, eg proportional hazard, delayed effect and diminishing effect. The weight on each scenario should reflect the prior belief and a reasonable approach at the design stage would be to optimize the corresponding assurance under the constraint of control of the type I error rate. 

Just like rMW, the MaxCombo also performs well with respect to power and assurance, for similar reasons as the rMW. However, as shown in Table~\ref{tabResultsHigh} and extensively discussed in \cite{magirr2019modestly}, it does not control the type~I error in the case of early harm. The original MaxCombo also included a weighted test that emphasizes early differences. We choose not to include an early-weighted test because our emphasis was on being robust to worse-case scenarios relative to the log-rank test when we expect a delay. Furthermore, some would argue that differences in survival at the end of follow-up are often (not always) of more importance. Nevertheless, if a diminishing effect is considered a strong possibility at the design stage, then including the early weighted test will, of course, increase power. 

We have focused on the analysis of a single primary endpoint but the approach can be readily extended to test multiple endpoints. If PFS is tested with a rMW, we can proceed to test OS if the null hypothesis for  PFS is rejected. The choice of test for OS could be based on prior assumptions just like for PFS, considering rMW and other candidate tests included in our simulation study for the primary endpoint. It would in principle not be a problem to for example use rMW for PFS and LR for OS, or vice versa. If an endpoint is to be tested across more than one interim analysis and final analysis within a group sequential framework, as is typically the case for OS and sometimes also for PFS in oncology trials, the methods used in \cite{ghosh2022robust} can be applied. 

%Testing and analysis, I amde an attempt at describing your's and Caffe's proposal as discussed at the June EMA meeting, does it make sense and is there a reference? 

Our focus has been on developing tests for the primary null hypothesis that while controlling the type I error rate, can provide robust power across a range of scenarios. Once the primary null hypothesis has been rejected, an important additional step would be to assess the magnitude of the treatment benefit. We believe that optimizing the testing is fundamental, as rejection of the primary null hypothesis is typically a necessary condition to support regulatory approval. That being said, we certainly acknowledge the importance of discussion and pre-specification of the target estimand and appropriate methods of estimation at the design stage, in addition to the choice of test for the primary null hypothesis. 

%\begin{itemize}
%\item Power, what do we gain/lose? Main selling point: can gain a lot when there is delayed separation but little if there is no delay or an early effect (modestly weighted alone would not be as robust)
%\item Weighted versions of tests: a rather simple formulation could be: what is the optimal $\alpha$ to allocate to the modestly-weighted log-rank test vs the standard test? Even with a very small penalty for the $\alpha$ allocated to the standard test, we could perhaps increase power/assurance quite a bit?  
%\item Consider uncertainty across reasonable scenarios, average power/(conditional) assurance? 
%\item Efficiency gains 
%\end{itemize}

\section*{Acknowledgments}

%Bibliography
\bibliographystyle{unsrt}  
\bibliography{references}

\end{document}